\documentclass[nhess, manuscript]{copernicus}

\begin{document}
 \nolinenumbers

\title{InSAR-Informed In-Situ Monitoring for Deep-Seated Landslides: Insights from El Forn (Andorra)}


\Author[1]{Rachael}{Lau}
\Author[1]{Carolina}{Segu\'{\i}}
\Author[1]{Tyler}{Waterman}
\Author[1]{Nathaniel}{Chaney}
\Author[1]{Manolis}{Veveakis}

\affil[1]{Department of Civil and Environmental Engineering, Duke University, Durham, USA.}




\correspondence{Rachael Lau (rachael.lau@duke.edu)}

\runningtitle{InSAR-Guided Monitoring of Deep-Seated Landslides in El Forn, Andorra}

\runningauthor{Lau et al}

\received{}
\pubdiscuss{} 
\revised{}
\accepted{}
\published{}


\firstpage{1}

\maketitle

\begin{abstract}
Monitoring deep-seated landslides via borehole instrumentation can be an expensive and labor-intensive task. This work focuses on assessing the fidelity of Interferometric Synthetic Aperture Radar (InSAR) as it relates to subsurface ground motion monitoring, as well as understanding uncertainty in modeling active landslide displacement for the case study of the in-situ monitored El Forn deep-seated landslide in Canillo, Andorra. We used the available Sentinel-1 data to create a velocity map from deformation time series from 2019-2021.  We compared the performances of InSAR data from the recently launched European Ground Motion Service (EGMS) platform and the ASF On Demand InSAR processing tools in a time series comparison of displacement in the direction of landslide motion with in-situ borehole-based measurements from 2019-2021, suggesting that ground motion detected through InSAR can be used in tandem with field monitoring to provide optimal information with minimum in-situ deployment. While identification of active landslides may be possible via the use of the high-accuracy data processed through the EGMS platform, the intents and purposes of this work are in assessment of InSAR as a monitoring tool. Based on that, geospatial interpolation with statistical analysis was conducted to better understand the necessary number of \textit{in-situ} observations needed to lower error on a remote-sensing recreation of ground motion over the entirety of a landslide, suggesting between 20-25 total observations provides the optimal normalized root mean squared error for an ordinarily-kriged model of the El Forn landslide surface. 
\end{abstract}


\introduction  
Deep-seated landslides represent one of the most devastating natural hazards on earth, many creeping at inappreciable velocities over several years \cite{Smalley1978,Voight1988} before suddenly collapsing, usually with catastrophic velocities. While there are a range of landslide sizes, several deep-seated landslides include  sizable earth slides involving millions of cubic meters of soil moving as a rigid block on top of a deep (below the roots of the trees and the groundwater level) basal layer of heavily deformed minerals \cite{petley1998,Frattini2013}. 
Their collapse is usually very sudden, happening within minutes and without a clear warning  \cite{Voight1988,Smalley1978}, reaching high velocities, as high as the 20 m/s reported at the 1963 Vaiont landslide in Italy \cite{Reid1994,Veveakis2007}. The catastrophic and fast collapse of this kind of landslides makes the evacuation of the area that could be affected a cumbersome task, thereby increasing risk of fatalities and infrastructure damages \cite{Reid1994,usgs,huang, Guzzetti2000}. Moreover, the complex physical nature of the landslides induces high uncertainty in the number of insitu observations required for a high-fidelity monitoring system. That, in combination with the challenging and expensive methods of insitu monitoring, makes the development of reliable, data-driven, early warning systems (or tools/protocols to stop the acceleration of the landslide) an appealing proposition.

Before the use of satellites, initial approaches in predicting the catastrophic collapse of a landslide rely on physical access in the area with in-situ (extensiometer) or ex-situ (LiDAR, UAV) displacement data \cite{jaboyedoff,yaprak}, whereby an assessment is made by using the inverse velocity method \cite{Voight1988,Saito,Carla2017,Zhou2020}. Considerable work has since been done in developing remote sensing methods for landslide identification \cite{Handwerger2019,Zhong2020,Mohan2020,Casagli2023,Chi2002, Zhao2018} as well as creating predictive models of deep-seated landslides based on identifying different mechanisms involved as triggering factors of the acceleration like rainfall \cite{Reid1994}, temperature \cite{Mitchell1968,Veveakis2007} and chemical alterations \cite{Hueckel2002}. Both developments are now at a stage were they can be used in conjunction with high fidelity field data (piezometers, extensiomenters and thermometers) to obtain forecasting and mitigation protocols \cite{Segui1,Segui2022}. However, the installation of such in-situ instrumentation is a costly operation, requiring the transportation of heavy equipment often in remote areas and the installation of sensors in deep boreholes, that cannot be deployed readily across the world.

To overcome these constraints, the use of remote sensing has become a more available tool for  landslide monitoring over the last several decades. Several techniques for mapping and assessing slope movements have been developed, thus allowing for more reliable and fast investigation \cite{Cigna2013,Fiorucci2011,Guzzetti2009,Michoud2012}. Among the remote sensing options, the use of Synthetic Aperture Radar (SAR) sensors has gained significant popularity for measuring surface deformations and constructing their time series, since this approach requires no access to the site to install borehole instrumentation or handle UAV and LiDAR devices.  Remote monitoring approaches for deep-seated landslide are limited by their often inability to provide information for the body of the landslide when the moving mass is deep-seated in steep valleys or densely vegetated mountain ranges, as well as their nature as surface-only measurements. This work builds on the existing literature concerning the assessment of how reliable remote surface measurement tools could be for deep-seated landslides \cite{Bayer2017,Fobert2021,Bellotti2014, Casagli2023,Sarkar2004,Lissak2020, Scaioni2014, Wang2019}, providing a case study from the El Forn landslide in terms of the data quality needed to identify and monitor a landslide and extending this body of literature by using InSAR data to decide the minimum number of in-situ observations needed for that.

\section{Material and Methods}

\subsection{Description of the El Forn Landslide and In-Situ Data}
\begin{figure}[h]
    \centering
    \includegraphics[width=0.8\textwidth]{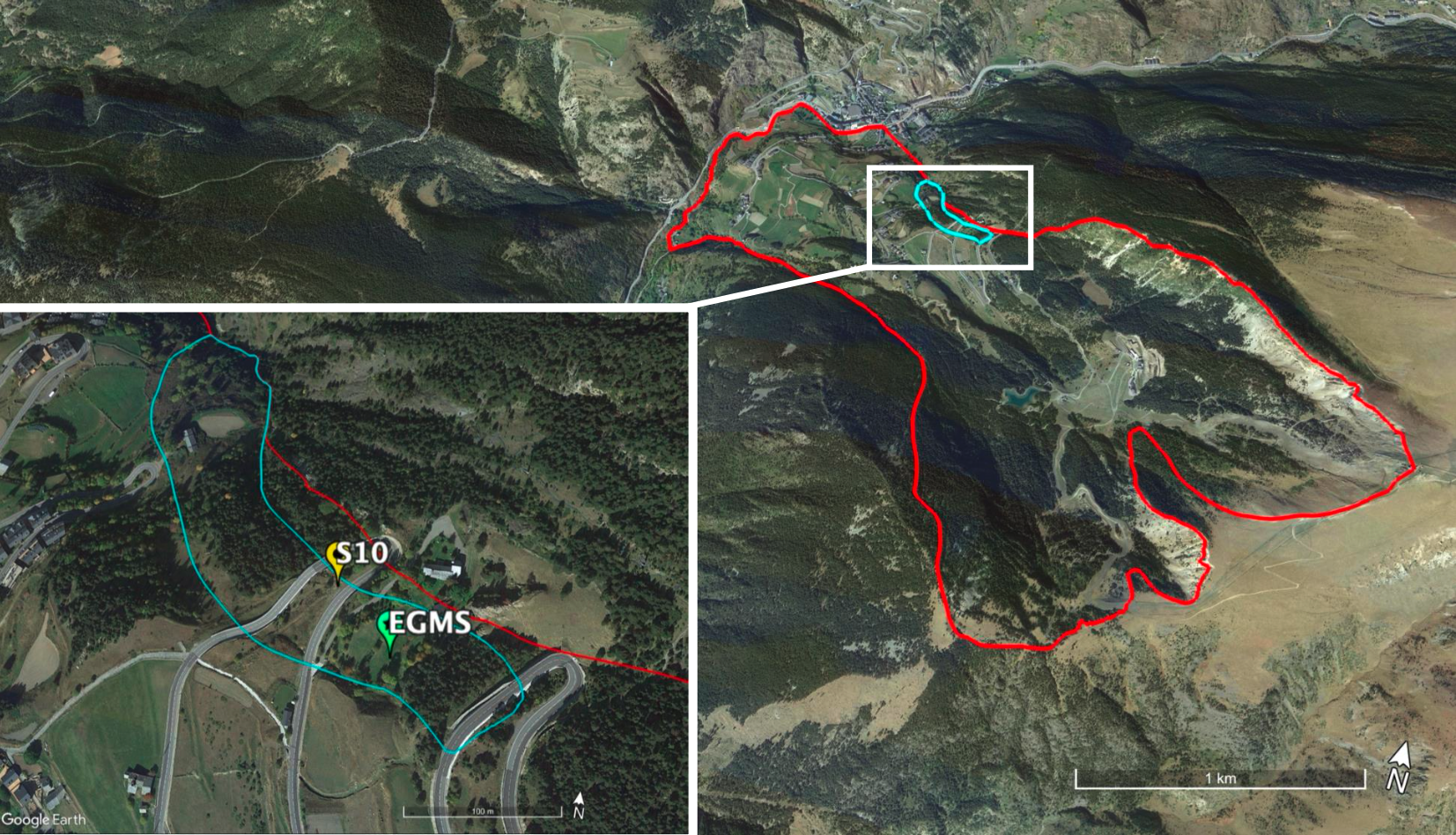}
    \caption{Overview of El Forn landslide with Cal Ponet-Cal Borronet lobe, noted with EGMS observation (see section 2.2.2) and S10 borehole location. Retrieved from \textit{Google Earth}, 2023.}
    \label{fig:f01}
\end{figure}

The El Forn landslide is a large deep-seated landslide located southeast of the town of Canillo, Andorra, nestled in the Pyrenees (see Figure \ref{fig:f01}) that is triggered by snow melt and season rainfall that collect into an aquifer located below the sliding surface. This landslide has a sliding mass of approximately 300 $Mm^3$ that creeps at an average rate of 1.2 cm/year \cite{Segui1}. Within the main sliding mass of the landslide, there is a faster-moving lobe (Cal Ponet-Cal Borronet lobe) that slides at a maximum velocity of 2-4 cm/year. At present, this lobe is equipped with 12 boreholes dispersed between the top and bottom of the landslide collecting continuous in-situ data. 

The sliding surface is located at 29m depth, and the landslide is moving as a rigid block \cite{Segui1} on top of it. The terrain of the landslide can be seen in Figure \ref{fig:f01}.

The primary instrumentation and data considered in this study are housed within borehole S10, noted as the yellow marker in figure \ref{fig:f01}. Data from S10 is sampled continuously every 20 minutes via instrumentation including an extensometer, three piezometers, and a thermometer within the shear band, which measure horizontal displacement, water pressure, and temperature changes in the material, respectively. The data considered in this study for El Forn are displacement gathered from the extensometer.

\subsection{Remote Data Collection and Processing}
One of the key objectives of this work is to compare InSAR to subsurface ground measurements. This is achieved through interferograms obtained by Sentinel-1 A/B over a period of 6 months in 2019 with a 6 day aquisition interval. The interferograms are then processed to obtain displacement time series over the landslide's surface using two different approaches:

(1) a high-precision (fine spatial resolution), low-accuracy (noisy) approach whereby Sentinel-1 data is retrieved and pre-processed with a low coherence threshold to obtain high spatial resolution (40x40 meter grid) displacement data so that geospatial analysis can be conducted to determine the minimum number and location of observations required for landslide monitoring and reconstruction with quantified uncertainty. This approach was deployed using the Alaska Satellite Facility's (ASF) Vertex Platform's On Demand InSAR processing tools and will be hereinafter reffered to as ASF; and

(2) a low-precision (sparse spatial resolution), high-accuracy (de-noised) approach whereby Sentinel-1 data are filtered to reduce the noise so that landslide identification can be achieved from high-accuracy data on a 100x100 meter grid. This approach was performed via immediate download through the newly-launched European Ground Motion Service (EGMS) Platform by Copernicus \cite{EGMS} and will be hereinafter referred to as EGMS. 

Note that the SAR imagery for both the data retrieved via the ASF On Demand InSAR processing tools and the Copernicus EGMS portal was taken from Sentinel-1 A/B satellites on a descending track with a 270-degree angle of incidence from the vertical. Using the slope of the ground at S10, the data for the EGMS displacement and ASF-MintPy readings were translated into the displacement along the direction of the landslide movement so it could be compared to S10’s strain gauge readings. 

While displacement data from EGMS is readily-available, data retrieval from ASF requires a more hands-on approach, going through a short baseline subset pre-processing step via ASF On Demand InSAR processing tools, followed by an interferogram time series inversion via the Miami InSAR time-series software for Python (MintPy) that allows us to generate mean deformation velocity maps and deformation time series \cite{Berardino2002,Handwerger2019,Yunjun2019}. Subsequent displacement data from this time series inversion, alongside displacement data pulled from the EGMS platform are compared with in-situ displacement data from S10 to understand correlation between InSAR and in-situ data. The other key objective of this work is to understand how InSAR can be used for general uncertainty quantification for planning future borehole placement, should the first objective prove InSAR can be correlated with sub-surface measurements. This will be done via iterative ordinary kriging, with the normalized root-mean-squared-error (RMSE) being used as the statistical parameter of interest for confidence. The next paragraphs outline the technical details of data retrieval and processing for each of these approaches.

\subsubsection{ASF InSAR Data Retrieval and Time Series Inversion} \label{sec:asf_retrieval} 

Open-access descending-track SAR acquisitions from Sentinel-1 C-band (5.66 cm radar wavelength) were pulled from the Alaska Satellite Facility’s (ASF) Vertex Portal and processed automatically through this portal via the Advanced Rapid Imaging Analysis (ARIA) for Natural Hazards Project \cite{Bekaert2019}. 

InSAR data retrieved for the purposes of this work were retrieved by selecting Single Look Complex (SLC) scenes with a beam mode of Interferometric Wide (IW) and cover the El Forn landslide. Using the Alaska Satellite Facility’s On Demand tool, scenes were selected and pre-processed using the Short Baseline Subset (SBAS) tool, making it easier to order the best interferograms for SBAS. However, MintPy's default time-series tool was ultimately used. From there,  all 619 interferograms covering El Forn were downloaded via Python script from the ASF Vertex Platform. Interferograms with visible discontinuities were manually identified once downloaded and removed from the stack for time-series analysis. Please see supplementary material S1 for more information on InSAR pre-processing and subsequent workflow. From there, the ASF Hybrid Pluggable Processing Pipelines (HyP3) service allowed for each interferogram to be clipped to the same size of overlap to standardize each interferogram. Using MintPy (see \cite{Yunjun2019} for more information on the time series inversion process), clipped interferograms were then inverted to create a deformation time series using a weighted least squares inversion with a coherence threshold value of 0.4. This approach creates velocity and deformation maps on a 40x40 meter grid, as shown in Figure \ref{fig:f02}(a). It is to be noted that the low coherence threshold used provides high spatial resolution maps that can be used for geospatial analysis, however this increased resolution is accompanied by increased noise, which makes landslide identification cumbersome, as seen in Figure \ref{fig:f02}(a) and discussed in the results section. For this reason, a second approach is pursued in parallel, focusing on the accuracy of the data as detailed below.

\subsubsection{EGMS InSAR Data Retrieval and Time Series Inversion}
A second set of InSAR data was taken from the European Union’s Copernicus project via the European Ground Motion Service (EGMS) portal, available for immediate retrieval as vertical and East-West displacement series per point. This platform provides already-processed displacement data over parts of El Forn at a grid of around 100x100 meter resolution as seen in Figure \ref{fig:f02}(b). 

As already mentioned, there are key differences between the data retrieved via the ASF On Demand InSAR processing tools and the data retrieved from the EGMS portal for the intents and purposes of this work -- the  key trade-off being between precision and accuracy. More specifically, the ASF On Demand data inverted via MintPy used a minimum threshold coherence value of 0.4, whereas EGMS Ortho Data only visualized individual measurement points greater or equal to 0.8. It is to be noted that the EGMS results could have also been obtained by applying a higher threshold value to the ASF approach, making the choice between the tool used scientifically immaterial. The reasons for utilizing both in parallel are the ability to showcase and cross-validate the two approaches.  

\subsection{Spatial Interpolation and Ordinary Kriging}

Ordinary kriging was conducted by first creating a grid of $x$- and $y$- coordinates and corresponding velocity values at these points. Distances between the random observations and each individual grid point were calculated, such that:

\begin{equation} \label{eq:1}
    d_1 = \sqrt{(x_g - x_{obs}^T)^2 + (y_g - y_{obs}^T)^2},
\end{equation}

where $x_g$ and $y_g$ are the grid coordinates, and $x_{obs}$ and $y_{obs}$ are the random observation coordinates. The covariance matrix were determined using the range $\tau$ and variance $\sigma^2$ from the semivariogram, such that:

\begin{equation}\label{eq:2}
 C = \sigma^2(e^{(-d_1/\tau)^T})
\end{equation} 

The euclidean distances between the random observations and each other were calculated, as well as the corresponding covariance matrix $\Sigma$, such that:

\begin{equation} \label{eq:3}
    d_2 = \sqrt{(x_{obs} - x_{obs}^T)^2 + (y_{obs} - y_{obs}^T)^2}
\end{equation}

\begin{equation}\label{eq:4}
    \Sigma = \sigma^2(e^{(-d_2/\tau)^T})
\end{equation}

The covariance matrices were appended into two matrices that would be used Lagrange Multipliers, into matrices $\Sigma'$ and $C'$, respectively. The weights were calculated by solving the linear equations created by the $\Sigma$ and $C$ matrices. From there, we calculate predictions $Z^*$ by taking the values velocity values at the random observations, $z_t$, multiplying them by the corresponding weights $W$, such that:
\begin{equation}\label{eq:5}
    Z^* = \Sigma(W*z_t)
\end{equation}

The mean squared error is then solved, such that:
\begin{equation}\label{eq:6}
    MSE = \sigma^2 - \Sigma(W*C')-W
\end{equation}

As a result, fidelity was assessed via root mean-squared error ($RMSE = \sqrt{MSE}$) of a kriged landslide surface done via random sampling done without replacement per iteration. 

\section{Results}
\subsection{Landslide Identification}
\begin{figure}[h]
    \centering
    \includegraphics[width=0.8\textwidth]{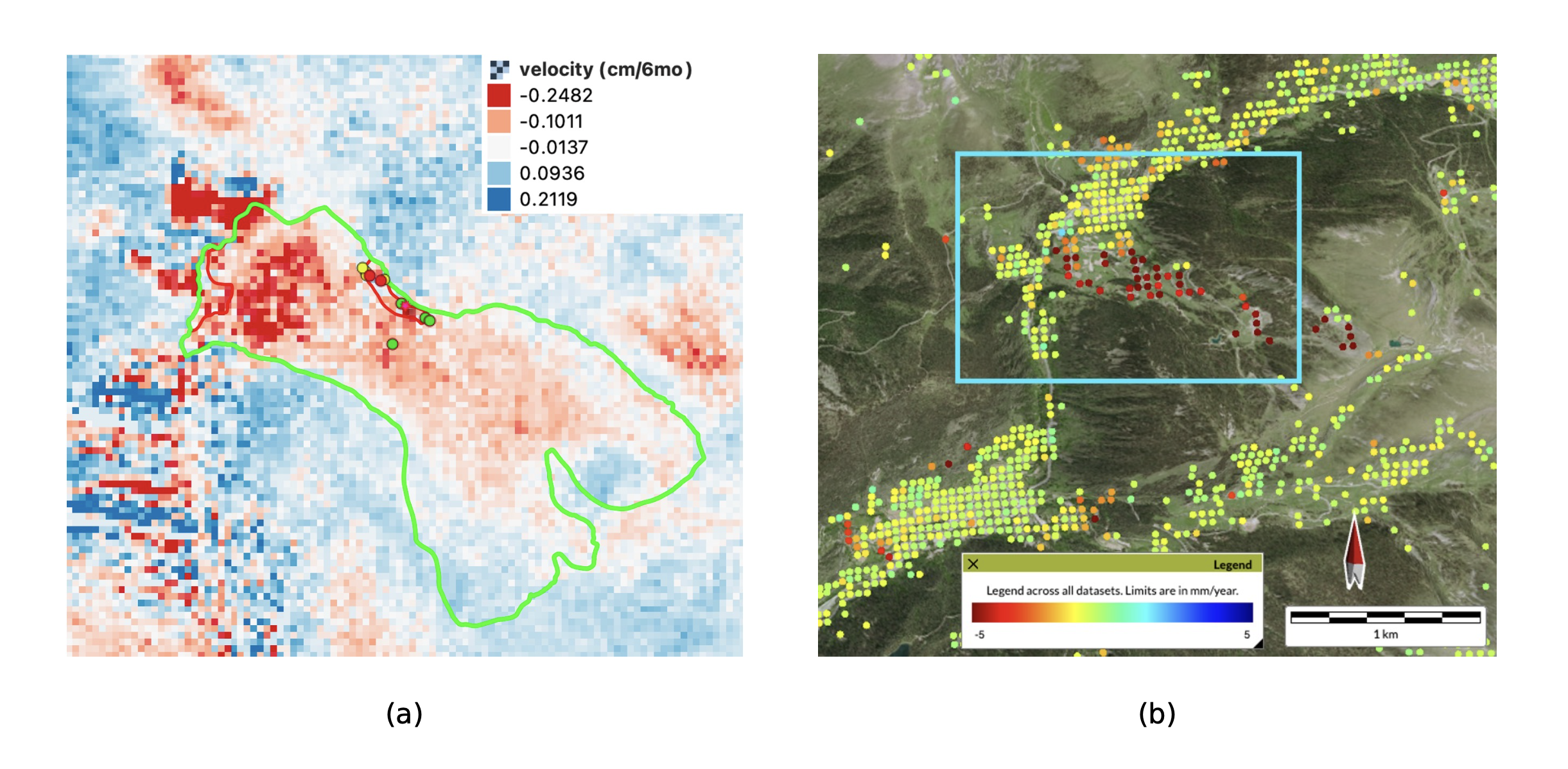}
    \caption{Depiction of El Forn landslide using InSAR. (a) Overview of El Forn landslide velocity using data retrieved and inverted using ASF against the field observation of the landslide boundary (green line). (b) InSAR detection of El Forn landslide by Copernicus EGMS platform (highlighted in blue), retrieved 28 September 2023. Indicative of possible use of EGMS as tool for active landslide detection.}
    \label{fig:f02}
\end{figure}

As previously noted, ASF and EGMS data showcase a key trade-off between precision and accuracy for the purposes of landslide monitoring. Figure \ref{fig:f02}(a) demonstrates that with a lower coherence value, more data is available, albeit noisy. In a separate vein, the EGMS data (pictured in Figure \ref{fig:f02}(b) demonstrates the utility of an increased coherence value in reducing noise and producing high-accuracy data usable for landslide detection.

\subsection{Correlating InSAR with In-Situ Data for seasonal ground motion}
\begin{figure}[h]
    \centering
    \includegraphics[width=0.8\textwidth]{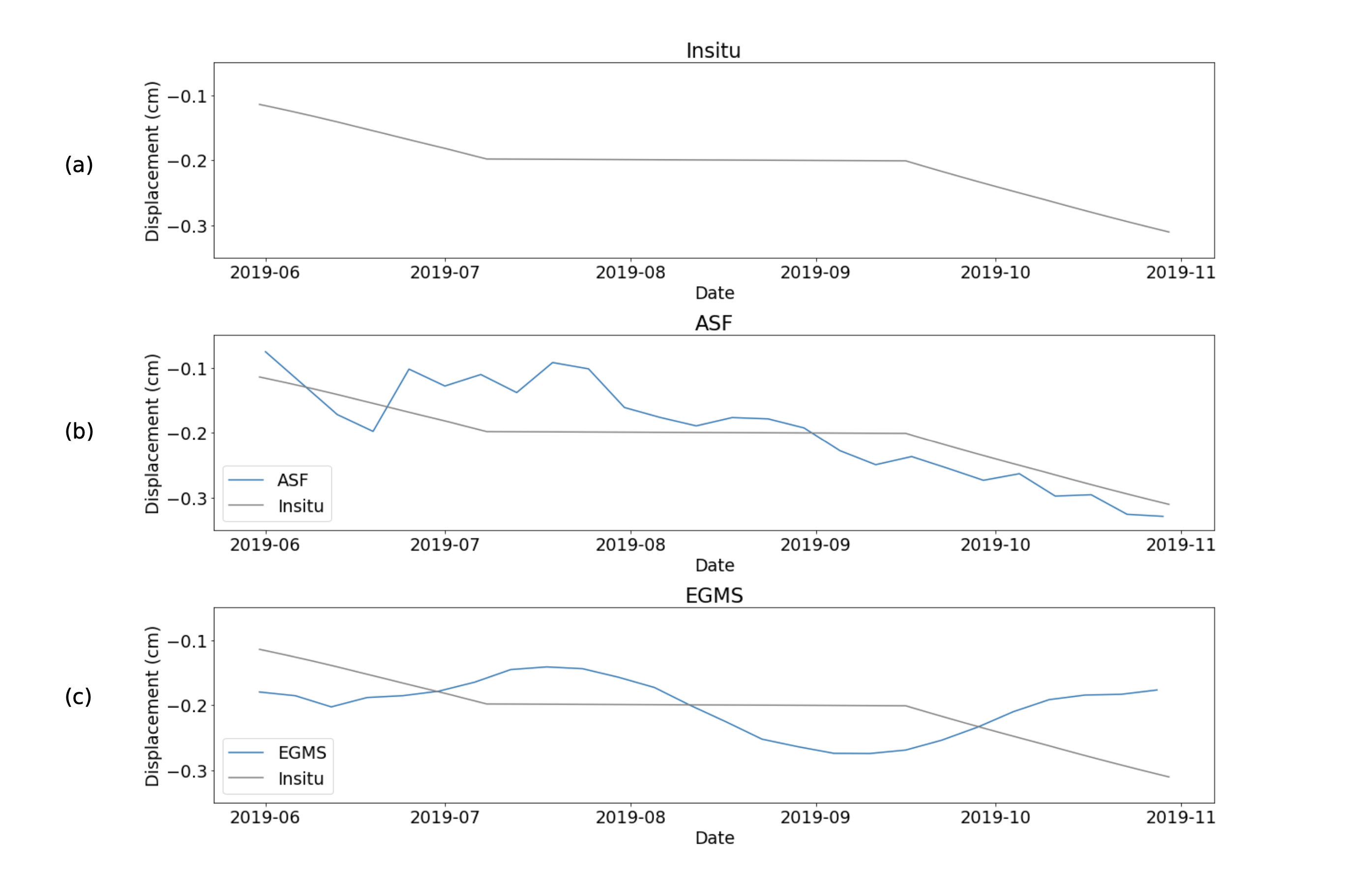}
    \caption{Comparison of insitu displacement data with displacement data retrieved via EGMS and ASF On Demand Processing tools. (a) Insitu displacement readings from S10 borehole. (b) 7-day cumulative moving average of InSAR displacement readings over S10 with data retrieved via ASF On Demand Processing tools. (c) 7-day cumulative moving average of InSAR displacement readings with data retrieved from EGMS.}
    \label{fig:f03}
\end{figure}

Upon data retrieval, both ASF and EGMS InSAR displacements were compared along the direction of sliding with insitu strain gauge data from borehole S10 in order to understand the fidelity of InSAR in monitoring sub-surface ground motion. This direct comparison of InSAR readings from ASF and EGMS over the S10 borehole can be seen in Figure \ref{fig:f03}. Retrieval of InSAR displacement data from EGMS required manual comparison of a couple of neighboring points with S10's insitu data in order to find points with a strong enough signal to use to compare since there was no individual measurement point at the location of S10 after the increased threshold was applied (see Figure \ref{fig:f01}).

Indeed, the increased sparsity from EGMS resulted in a lack of precision of the individual measurement points to compare with insitu measurements, as seen in Figure \ref{fig:f03}(c), as compared to data retrieved via ASF's On Demand tools (as seen in Figure \ref{fig:f03}(b). Since data for the exact location of S10 borehole on the landslide  was not immediately available on the EGMS platform, two coordinates neighboring the WGS-84 coordinate of S10 were pulled and compared to the S10 data and InSAR data. Heterogeneity within the landslide prevented selecting just one point as close to the S10 point as possible, without properly examining other neighbors. Figure \ref{fig:f01} details which point was examined, with “EGMS” being the point in the EGMS database that was ultimately used because of its closest alignment with S10’s raw displacement data. Figure \ref{fig:f03} directly compares data retrieved with ASF On Demand Processing tools (and inverted via MintPy) and EGMS with insitu displacement measurements. We observe that while InSAR displacement measurements pulled from EGMS are helpful in detection, the higher accuracy creates a lack of precision necessary for insitu comparison. 

In order to justify this claim and quantify the performance of the two approaches in time, a measure of linear independence (correlation) was conducted with data from EGMS and ASF with insitu measurements, respectively. The equation used for the calculation of the correlation coefficient $\rho(A,B)$ of two datasets $A$ (in this case EGMS or ASF) and $B$ (in this case the insitu data)  is:

\begin{equation} \label{eq:0}
    \rho(A,B) = \frac{1}{N-1}\sum_{i=1}^{N}\left(\frac{A_i -\mu_A}{\sigma_A}\right)\left(\frac{B_i-\mu_B}{\sigma_B}\right),
\end{equation}
 
where $\mu$ and $\sigma$ are the mean and standard deviation of the data sets, respectively. The results of the performance of ASF and EGMS against the insitu data are detailed in Table \ref{table:tabel1}, where we see that the ASF dataset ($\rho(ASF,InSitu) = 0.6957$) has a considerably better performance against insitu data than EGMS ($\rho(EGMS,InSitu) = 0.0761$). This is presumably due to the spatial heterogeneity of the landslide's displacement around the point of measurement, which is not exactly on S10 (see Figure \ref{fig:f01}), and is not reflective of the overall quality of the EGMS data. 

\begin{table}[!htb]
 \centering
 \caption{Comparison of Correlation Coefficients (Eq.\ref{eq:0}) of displacement data retrieved from ASF and EGMS with insitu measurements.}
    \begin{minipage}{.5\linewidth}
      \centering
        \begin{tabular}{llll}
         &\vline \: InSitu & ASF & EGMS \\ \hline
        InSitu &  \vline \: 1.000 & 0.6957 & 0.0761  \\
        ASF &  \vline \: 0.6957 & 1.000 & -  \\
        EGMS &  \vline \: 0.0761 & - & 1.000
         \end{tabular}
    \end{minipage}%
\label{table:tabel1}
\end{table}

\subsection{Ordinary kriging: determining necessary number of remote observations}
\begin{figure}[h!]
    \centering
    \includegraphics[width=1.0\textwidth]{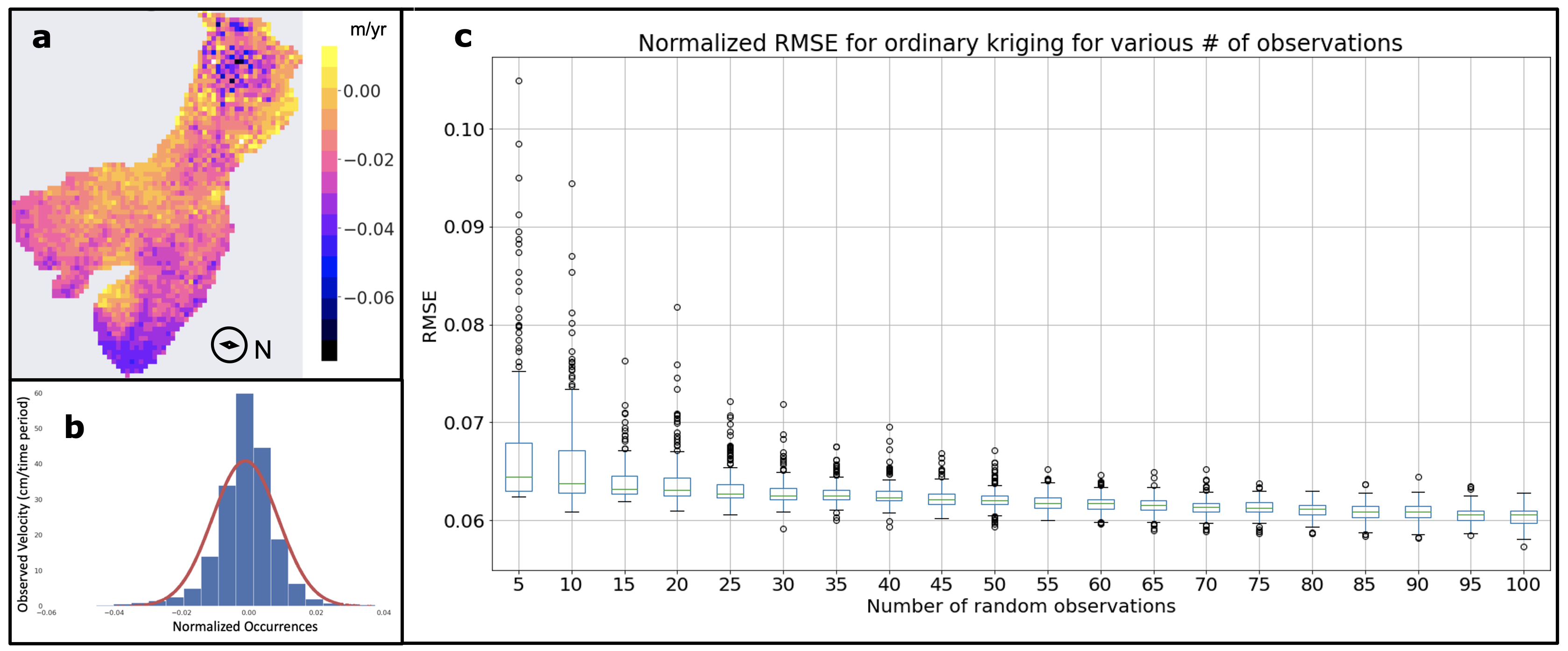}
    \caption{(a) InSAR velocity map of the 2019 snow-free period in Andorra via ASF-processed InSAR, (b) uniform distribution probability density function (red line) and occurrence histogram  (blue) of velocities pulled from sub-figure (a), (c) boxplot of normalized RMSE of 200 iterations for various number of random observations of velocities from sub-figure (a) pulled from uniform distribution in sub-figure (b).}
    \label{fig:f04}
\end{figure}

Having shown the correlation between ASF and insitu in the previous section, we move forward with the densely-populated ASF mapping to carry out ordinary kriging. In order to best understand how many observations (i.e., boreholes) impacts the ability to remotely model ground motion of a deep-seated landslide, 200 iterations of randomly-selected samples (with sizes ranging from 5-100 points) along the main landslide surface were selected and had ordinary kriging performed on them to assess the RMSE in predicting ground motion over the surface of the landslide, with summaries of RMSE for each number of iterations visible in the box and whisker plot in Figure \ref{fig:f04}. Similarly, single-iteration ordinary kriging was conducted over the sliding mass to assess how various sample sizes (ranging from 10-2000) were in \textit{recreating} velocities of the sliding mass, and where possible areas of interest for further investigation were. 

More specifically, 200 iterations were conducted per number of random observations of the average velocity in 2019 (pulled from a uniform distribution, as seen in Figure \ref{fig:f04}b), results of which can be seen in Figures  \ref{fig:f04}c. Figure \ref{fig:f04}c, the box for the normalized root mean squared error (RMSE), indicates a marked drop in the interquartile range (IQR) between 20 and 25 observations. Also important to notice is that the range of outliers is significantly lower starting from n = 25 observations and forward. Note that the dots outside of the whiskers are outliers, meaning they lie outside of the whiskers defined by $Q1 - 1.5 * IQR$ and $Q3 + 1.5 * IQR$. 

\begin{figure}[h!]
    \centering
    \includegraphics[width=0.8\textwidth]{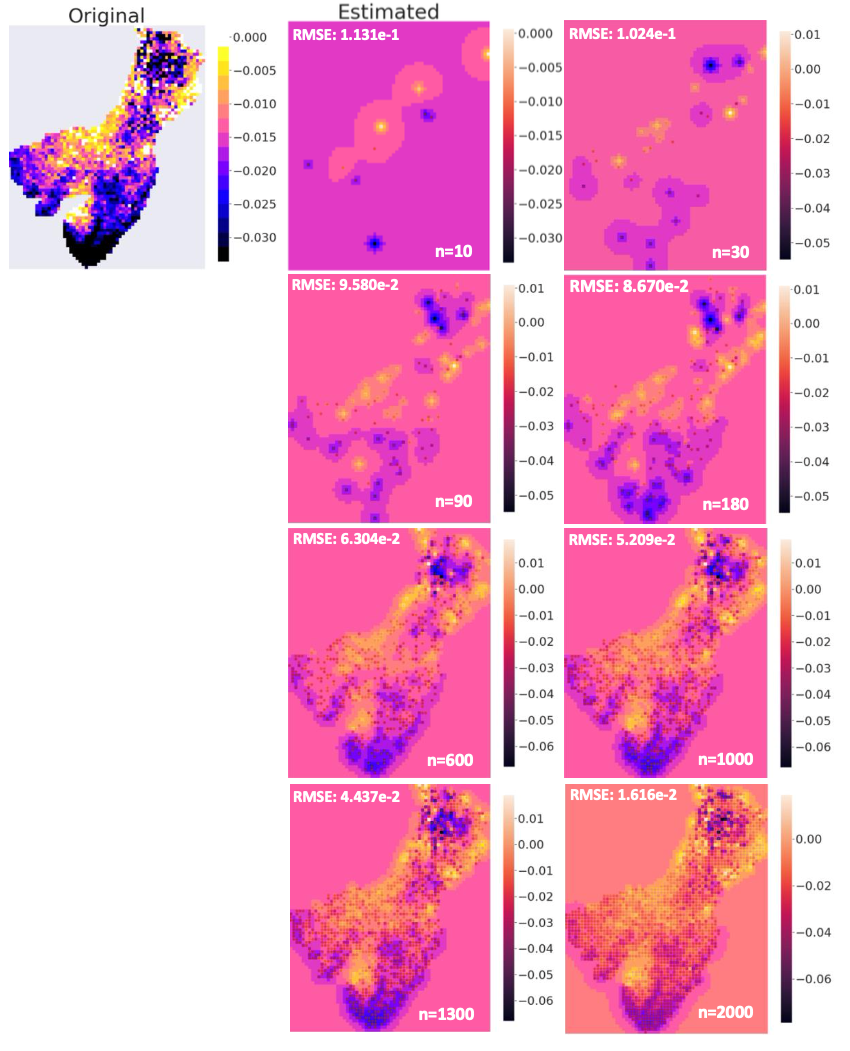}
    \caption{Ordinary kriging results of various random samples (n = 10 - 2000) via one iteration (as opposed to the 200 iterations in Figure 3(c)), reflecting gaps in predictive capabilities on the surface of the landslide for further investigation. Normalized RMSE for each ordinary kriging process indicating  error for each sample size.}
    \label{fig:f05}
\end{figure}

Figure \ref{fig:f05} reflects how \textit{n} samples recreates the landslide surface movement and the fidelity (RMSE) of doing so. For example, n = 30 random samples recreates certain parts of the landslide better than others for one iteration. Figure \ref{fig:f05} shows the evolution of how increased random samples that go through ordinary kriging process then recreate certain parts of the landslide surface faster or better than others. In the case of n = 30, the top and bottom of the landslide are better developed than the middle of the landslide, which indicates where further investigation may be necessary. More specifically, the center of the landslide is the least developed throughout the ordinary kriging iterations -- for modeling purposes then, further investigation would be required on this part of the sliding mass (either further instrumentation or a more narrow scope of InSAR). 

\conclusions  
In this paper, the use of InSAR for landslide monitoring was assessed for two key objectives: (1) correlation with insitu data to test the accuracy of InSAR in monitoring seasonal and off-seasonal sub-surface movement, and (2) spacial interpolation across the landslide surface with various number of InSAR points to help us understand use of InSAR for establishing areas on a scarp that need monitoring (i.e. further instrumentation), as well as understanding how many remote observations would allow us to minimize error in recreating the scarp without using the full data set. Correlation of the InSAR data with extensometer data in S10 borehole on the El Forn scarp indicates that InSAR can be used to understand seasonal sub-surface ground motion. 

The spatial interpolation, as well as the susbequent error assessment, conducted on the El Forn landslide using solely InSAR data helped determine the necessary number of observations to adequately monitor the general movement of the landslide. Based off of 200 iterations of random samples going through an ordinary kriging process on the landslide, the outliers of the normalized root mean squared error dropped significantly between 20 and 25 remote observations, as indicated in Figure \ref{fig:f04}. Based off of Figure \ref{fig:f05}, the most uncertainty, coupled with the most movement, through even an increased number of random samples is in the middle of the landslide, can be seen in the middle of the top left lobe, in the northeast corner of the landslide. In future studies, we could look to perform regression kriging with 20-25 remote observations, focused solely in this region to understand how uncertainty propagates for this part of the landslide on a finer time scale. Overall, InSAR has many purposes when considering the monitoring of deep-seated landslides, with several options to build on our existing knowledge for studies to come.



\dataavailability{Data sets are open-access and retrievable from the Alaska Satellite Facility Vertex Platform (https://asf.alaska.edu/) and the European Ground Motion Service Copernicus Platform (https://egms.land.copernicus.eu/).} 


\sampleavailability{Insitu sample data is available upon request of the corresponding author.} 


\appendix
\section{Alaska Satellite Facility (ASF) InSAR Workflow}    
The Alaska Satellite Facility’s Vertex Platform user-friendly interface allows for ease of specifications on selecting an InSAR pair. For this work, an interferometric-Wide Single Look Complex (IW SLC) pair was selected – SLC meaning that the SAR data has been compiled to an image but hasn’t been multi-looked yet. Once the reference and dates of interest are selected, ASF begins multi-looking through various pairs of images. Note that for continuity purposes, the older SLC image is always used as the reference image.

In order to prepare a digital elevation model (DEM) file for subsequent geocoding and corrections, a topographic phase is subtracted from the interferogram by replicating an existing DEM to account for the actual topographic phase. In this case, Hyp3 takes the DEM from the publicly-available 2021 Release of Copernicus GLO-30 DEM library. Removing this topographic phase from the interferogram, the deformation signal is all that remains \cite{hyp3}.

Left with a stack of wrapped interferograms, phase unwrapping uses a Minimum Cost Flow (MCF) triangulation method to assign multiples of 2*pi to each pixel, which restricts the number of 2*pi jumps in phases to regions where they may occur. Note that thermal noise and interferometric decorrelation can result in 2*pi phase discontinuities, which are known as “residues” – these can be reduced via filtering. Filtering reduces phase noise and increases the accuracy of interferometric phase by reducing the number of interferogram residues\cite{hyp3}.
 
After filter, a validity mask directs the unwrapping process by applying thresholds for coherence and amplitude (backscatter intensity) values for each image pair. For this work, this amplitude threshold is kept to 0.0, so coherence thresholds drive the masks. Coherence is estimated from a normalized interferogram, with a range from 0.0 to 1.0, with 1.0 being perfectly coherent. Once coherent thresholds are applied, unwrapping will proceed relative to a fixed pixel point – one that should have a fixed pixel point. For this work, this point was selected as a rooftop in the town of Canillo at the foot of the landslide. This reference point is assigned an unwrapped phase value of 0 at this point, and every other pixel around it is then assigned a multiple of 2$\pi$ with respect to that point.
 
Lastly, these pixels are reprojected from SAR slant range space into a map-projected ground range-space and exported from the GAMMA internal format to GeoTIFF format. These unwrapped interferograms are ready to be go through a time series inversion\cite{hyp3}.


\noappendix       




\appendixfigures  

\appendixtables   


\authorcontribution{R. Lau conceived, designed, and carried out the analysis, as well as wrote the entirety of the manuscript. Under the Graduate Student Training Enhnacement Grant at Duke University, R. Lau worked under A. Handwerger where he provided critical resource guidance on the use of InSAR, ISCE+, and MintPy for this work. C. Segui provided helpful information regarding geophysical modeling and geology of the El Forn landslide. T. Waterman provided a helpful resource in getting set up and trouble shooting working on the computing cluster. N. Chaney provided space on his lab's computing cluster, as well as helpful guidance on spatial data analysis.  M. Veveakis provided supervision and writing guidance throughout the entirety of the research process.} 

\competinginterests{The author(s) declare(s) that there is no conflict of interest regarding the publication of this article.
} 


\begin{acknowledgements}
We acknowledge the collaborative efforts of our co-authors for their expertise and contributions to this work. Financial support from the National Science Foundation and the Fulbright Program was instrumental in conducting our investigations. We also appreciate the cooperation of the Government of Andorra in providing crucial in-situ data measurements. These combined efforts significantly enhanced the quality and scope of our study.
\end{acknowledgements}

\end{document}